\documentstyle[aps,prb,multicol,epsf,picinpar]{revtex}

\def\ep {\epsilon}
\def\ek {\epsilon_k}
\def\e2 {\epsilon-\epsilon_k}
\def\be {\begin{equation}}
\def\ee {\end{equation}}
\def\bea {\begin{eqnarray}}
\def\eea {\end{eqnarray}}
\def\om {\omega}

\begin{document}
\draft
\title{Metal-insulator transition in 2D: the role of interactions and
disorder}

\bigskip
\author{George Kastrinakis}
\address{ 
Institute of Electronic Structure and Laser (IESL), Foundation for
Research and Technology - Hellas (FORTH), \\
Iraklio, Crete 71110, Greece$^*$}
\address{
Dept. of Chemical Engineering, University of Cambridge,
Cambridge CB2 3RA, U.K. }

\date{September 7, 2004}

\maketitle
\begin{abstract}
We present a model for the metal-insulator transition
in 2D, observed in the recent years.
Our starting point consists of two ingredients only, which are ubiquitous
in the experiments: Coulomb interactions and weak disorder
spin-orbit scattering (coming from the interfaces of the heterostructures
in question). 
In a diagramatic approach, we predict the existence of a
characteristic temperature $T_o=T_o(n,\omega_H)$, $n$ being the
density of carriers, and $\omega_H$ the Zeeman energy, below which 
these systems become metallic.
This is in very good agreement with experiments,
and corroborates the fact that varying $n$ and
$\omega_H$ are equivalent ways into/out of the metallic regime.
The conductivity, calculated as a function of temperature and $\om_H$
in the metallic state, compares favorably to experiment. 
Moreover, we give an explicit expression for the conventional weak
disorder contributions to the conductivity in the frame of our model.
We comment on the nature of the transition, and calculate the specific
heat of the system.

\end{abstract}


In the last years, a metal-insulator transition has been
observed in 2D systems by Kravchenko et al. \cite{krav} and 
others - see e.g. \cite{rev}. In high mobility heterostructures,
for carrier densities $n$ higher than a critical density 
$n_c \sim 10^{11} cm^{-2}$,
and at appropriately low temperatures T, of a few degrees Kelvin at {\em most},
a transition into a metallic state is observed \cite{rev}. Moreover,
a sufficient increase of the Zeeman energy $\omega_H$ of the carriers, 
through an externally 
applied magnetic field $H$, takes the system back into the insulating state
\cite{rev}. 
Hamilton et al. \cite{reen1} and Pudalov et al. \cite{reen2} have
shown that beyond a maximum characteristic $n_c'$ there is a second
metal to insulator transition.
Further, in the metallic regime, the resistivity is usually fitted
very well by the formula - e.g. \cite{pud,rev}:
\begin{equation}
\rho_{exp}(T) = \rho_o \; +\;\rho_1 \; \exp{[-(T_*/T)^k]} \;\;, \label{ex0}
\end{equation}
where $\rho_o,\rho_1$ and $T_*$ depend on $n$. The exponent $k$ is in
the range 0.5 - 1, and is material-dependent.

Ilani et al. \cite{yacoby} have shown that for $n<n_c$ the 
insulating state is spacially inhomogeneous. Here, we will not attempt 
to provide a description of the insulating state. We {\em only} give 
a mechanism 
for the transition and a description of the metallic state. These
are based on
strong spin-dependent particle-hole correlations, which arise for
the appropriate range of $n$ and $\omega_H$, in the frame of a Fermi
liquid formulation.

The experimental systems in question have two ubiquitous characteristics.
The ratio of the Coulomb energy to the Fermi energy $\epsilon_F$ is typically
in the range 4-40 \cite{rev}. The importance of interactions has been
decisively demonstrated by Ilani et al. \cite{yacoby}
and Dultz and Jiang \cite{jiang}, who probed the compressibility
in the metallic regime, and found it in agreement with a many-body interacting
picture. On the other hand, weak disorder spin-orbit scattering
is inadvertently present, coming from the interfaces of the
heterostructures in question. These two characteristics constitute
our starting point. We note that magnetic spin disorder would
have the {\em same} effect as spin-orbit disorder, in the frame of our model.
Strong support for the role of spin-dependent scattering is provided
by the expts. of Vitkalov et al. \cite{vitk1}, which show that spin
up and down mobilities remain comparable for all $H$. This is naturally 
interpreted in terms of strong spin up and down mixing through 
disorder scattering.  
Further, Ilani et al \cite{yac2} have demonstrated the existence of localized
charged islands in the metallic phase, which can be magnetic. 
If this is the case, we can assume
that these magnetic islands sit on top of the aforementioned
spin-orbit impurities. Then the scattering strength of the impurities
is enhanced and the metallic behavior is more pronounced - c.f. e.g.
eq. (6) for $T_o(n,\om_H)$ etc.

In the foregoing, we consider the coupling $U$ between opposite spin 
carriers, and we denote by
\begin{equation}
u\equiv U N_F \;\;,
\end{equation}
the dimensionless coupling, $N_F$ being the density of states at the
Fermi level. As usual, we work in the regime
$\ep_F \tau > 1$ ($\hbar=1$), 
$\ep_F$ being the Fermi energy and $\tau^{-1}$ the total
impurity scattering rate. As we have shown in \cite{gk},
in the presence of weak disorder, which includes {\em spin} scattering,
the ladder diagrams in the particle-hole channel
give rise to the propagators $A^j(q,\omega_m)$, $j=0,\pm1$,
obeying the coupled Bethe-Salpeter equations 
\bea
A^1=U + U {\cal D}^1 A^1 + U {\cal D}^0 A^0 \;,\;  \\
A^0= U{\cal D}^0 A^1 + U {\cal D}^{-1} A^0 \;.   
\label{ex1}
\nonumber
\eea
The variables $q,\omega_m$, which stand
for the momentum and Matsubara energy difference between particle 
and hole lines respectively, were supressed. 
${\cal D}^j$ are given by ${ \cal D}^{\pm1}={\cal D}^{1,\pm1},\;
{\cal D}^0=
[ {\cal D}^{0,0} - {\cal D}^{1,0}]/2,\;
{\cal D}^{j,m_j}(q,\omega_m)=N_F 
\{Dq^2+j4\tau_{S}^{-1}/3\}/
\{Dq^2+j4\tau_{S}^{-1}/3+|\omega_m|-im_j\omega_H \}$,
with $\tau_S^{-1}$ the total spin scattering rate and
$D$ the diffusion constant.
In Appendix A we present the derivation of these equations and their
solution.

What turns out to be of interest here, is the "dynamic limit"
$Dq^2 < \omega_m$, where the solution of these eqs. is \cite{gk}
\begin{equation}
A^j(q,\omega_m) = \frac{K_{u j}+L_{u j} Dq^2+M_{u j}|\om_m|}
{A_u Dq^2 + B_u |\omega_m| + C_u } \;,
\label{eqsol}
\end{equation}
with
\bea
A_u=12-20 u+15 u^2/2+6 \Omega_{H}^2 \;\;, \;\;
B_u=4-6 u+3 u^2/2+2 \Omega_{H}^2 \;\;, \\  
C_u=r [\;1-2 u+3 u^2/4+\Omega_{H}^2 (1-u^2/4)\;]  \;\;, \;\;
K_{u0}=U r u (1+\Omega_H^2)\;\;, \;\; K_{u1}=U r (1-u)\;\;, \nonumber \\
L_{u0}= 2 U u(2+\Omega_H^2)\;\;, \;\; L_{u1}=4 U (1-u)\;\;,\;\;
M_{u0}= 2 U u(2+\Omega_H^2)\;\;, \;\; M_{u1}=U(4-3 u) \;\;, \nonumber 
\eea
where $\Omega_H = 3\omega_H \tau_S/4$ and $r=4\tau_S^{-1}/3$ $\;$ \cite{prp}.

We consider first the case with $\omega_H=0$.
For $u=2/3 - 2$, $C_u<0$. For $u\geq 0.845$, $B_u\leq 0$ as well.
Then the ratio $C_u/B_u$ is {\em negative} for $u=2/3 - 0.845$, with
$C_u<0$ and $B_u>0$. 
We interpret this as the onset of strong particle-hole correlations
in the spin density channel.
For sufficiently low temperatures
\begin{equation}
T < T_o(u,\omega_H) \equiv \frac{|C_u|}{2 \pi B_u}
\equiv \frac{\om_o}{2 \pi}\;\;,  \label{ew0}
\end{equation}
a {\em resonance} occurs - c.f. also eq. (\ref{expo}) , 
which we interpret as driving the 
second-order transition
into the metallic regime - c.f. below. 
A finite $\omega_H$ shrinks the range of $u$,
which allows for this effect, until for high enough $\omega_H$
$C_u$ becomes positive definite, and there is {\em no} transition
into the metallic state.
The dependence of the transition temperature $T_o$
on $u$ - or equivalently the carrier density - and $\omega_H$  
are in accordance with experiment \cite{reen1,reen2,sk,rev}.
Namely, the lobe within which the conducting state exists in the
$\omega_H$ vs. density diagram \cite{reen1,reen2} is easily
reproduced by use of the ratio $C_u/B_u$. 
Of course, the value of $u$ is determined through the bandstructure by
the density $n$.

We note that additional insertions in the Bethe-Salpeter 
equations (\ref{ex1}) for $A^j$,
such as self-energy diagrams, do not influence
the existence of this resonance. Based on the experimental evidence
available to date, 
there is {\em no} magnetic or superconducting instability involved
here. Thus the location of the pole of $A^j$ on the imaginary energy 
axis is unphysical. 
In a different approach, Chamon et al. \cite{cmc} 
assume a paired state in the metallic phase.
In order to make progress, we assume that in the metallic phase 
the usual Fermi liquid picture continues to apply, but $C_u$ acquires
a small imaginary factor
\be
 C_u \rightarrow C_u+i d \;\;,
\ee
which moves the pole of $A^j$ at a distance off the imaginary axis,
thus yielding well-defined metallic contributions to the conductivity.
$d$ should be such that the contribution of the diagram B in fig. 1
is substantial for the calculation of the conductivity, as mentioned in
Appendix B. Namely the quantity (c.f. eqs. (\ref{rj1}) and (\ref{expo})
for $R_J$ and eqs. (\ref{ldif}) and (\ref{logc})
for $D(q,\om)$)
\be
Q_{jj'} = \max\{G_R(k,\ep) G_A(k,\ep)\} |R_j R_j'| \;
\{\text{coefficient of} \sum_q D(q,\om_T) \}
= \frac{(2\tau)^2 |R_j R_j'| }{8 \pi^2 \tau^2 N_F D} \;\;,\label{sy1}
\ee
should not be too small. Also, $d$ should satisfy
\be
d=\pm d_o |C_u|^\xi T^\chi \;\; , \label{sy2}
\ee
with $0\leq\xi\leq 1$  \cite{comtr}, 
$0\leq \chi\leq 1$ and $T^\chi d_o(T) \rightarrow 0$
for $T\rightarrow 0$, in order to yield 
physically meaningful contributions (the sign $\pm$ corresponds to the position
of $A^j$ with regards to the main particle-hole lines, as shown in figs. 1-2). 
In fact, a finite $d$ is consistent with the finite imaginary part
of the dephasing rate for {\em finite} $\omega$ - c.f. e.g. \cite{deph}.

To calculate the conductivity
using the renormalized $A^j$, we proceed as follows.
First, relevant processes must take into account the resonance of
$A^j(q,\om)$. For this, diagrams, such as the ones in figs. 1-2, 
containing
pairs of $A^j(q,\om)$ 's with identical $q$ and $\om$ are needed in
order to yield the enhancement factor $Q_{jj'}$ of (\ref{sy1}) above, which 
then appears in eq. (\ref{msi}). 
The diffusons \cite{aa} 
\be
D^{j,m_j}(q,\om_m)=\frac{1}{2\pi N_F \tau^2} \; \frac{1}
{Dq^2+j4\tau_{S}^{-1}/3+|\omega_m|-im_j\omega_H} \label{ldif} 
\ee
in between the $A^j$'s, both provide
essential $T$-dependence and separate the $A^j$ 's from each other.
We note that without something separating the 2 $A^j$ 's, they would
just reduce into a {\em single} $A^j$, and thus not offering this
enhancement factor. 
The diagrams shown in figs. 1-2 yield the maximum
contribution around a single diffuson $D(q,\omega_m)$. The reasoning here
is similar to a related, but different, calculation in \cite{gk}.

Summing over all possible spin combinations involving the different $A^j$
's yields for the total contribution of B
\be
B = g\{ z_1  F_1+z_2 F_2\}  \;\;.
\label{msi}
\ee
Here,
$g=\frac{N_F}{2048 D^3}
\left\{\frac{T \om_o}{\pi^3 A_u  \ep_F^2 d}\right\}^2 ,\;
z_1=(K_{u1}+M_{u1} \om_o)^4+(K_{u0}+M_{u0} \om_o)^4, \; 
z_2=2 (K_{u1}+M_{u1} \om_o)^2 (K_{u0}+M_{u0} \om_o)^2 \;,\;
F_1=\ln\left(\frac{2\pi T+T_1}{2\pi T}\right) \;\;, \;\;
F_2=\frac{1}{2} \ln\left(\frac{(2\pi T+T_1+r)^2+\om_H^2}
{(2 \pi T+r)^2+\om_H^2}\right) $ and $T_1=\tau^{-1}$.

Subsequently, we also take into account all the diagrams of the type of fig. 2,
with one or two impurity lines passing completely outside the bubbles 
containing $A^j$'s. Summing all diagrams from figs. 1 and 2 gives
\be
M = \Gamma_{1} - \Gamma \Gamma_{2} \;\;,
\ee
where $\Gamma = G_R(k,\ep_F) G_A(k,\ep_F)$, 
$\Gamma_{1} = \alpha \tau^2 B, 
\Gamma_{2} = B/(2 \pi \epsilon_F \tau)$,
with $\alpha = 1 - (2\tau)^2 \left(\frac{4}{9 \tau_S^2}+
\frac{1}{3 \tau_S } \left[\frac{1}{\tau}-\frac{1}{\tau_S}\right] \right)$.

To calculate the conductivity, taking into account $M$ to infinite order,
we sum the infinite
series, the n-th term of which contains n factors (blocks) $M$ in series,
sandwiched between the current vertices of a conductivity bubble :
\be
\sigma_M = \frac{2 e^2}{m^2}\int d\vec{k} \; k_x^2 \left\{ \Gamma \;
\sum_{n=1}^{\infty} \left( \Gamma M \right)^n = 
\frac{\Gamma^2 M}
{1-\Gamma M} \right\} \;\;.     \label{sm1}
\ee
$m$ is the carrier mass.
In this way we obtain a $M$-dependent formula for the total conductivity
in the metallic regime
\begin{equation}
\sigma = \sigma_o \; + \; \sigma_M   
= \frac{2 N_F e^2 \epsilon_F \pi}{3 m S}
 \left\{ \frac{y_+}{ \sqrt{t^2 - y_+} }
- \; \frac{y_-}{ \sqrt{ t^2 - y_-} }
\right\} \; s_\sigma(T)
\;\;. \label{sgma}
\end{equation}
Here $y_\pm = (\Gamma_{1} \pm S)/2, \;
S=\sqrt{\Gamma_{1}^2 - 4 \Gamma_{2}}$.
$\sigma_o$ is the Drude term. 
$s_\sigma(T)$ is a smooth analytic function, equal to 1 for $T \leq T_o$
and gradually vanishing for $T>T_o$ \cite{exps}.
Eq. (\ref{sgma}) is a gauge-invariant
approximation for the conductivity \cite{gk,gau}.

The fit of eq. (\ref{sgma}) to the activated exponential in $T$ form for the
resistivity of eq. (\ref{ex0}), usually fit to
the experimental data, is shown in fig. 3. We take $\chi=1$ - which gives 
an excellent match with eq. (\ref{ex0}) in the "high $T$" limit, and makes
the factor $g$ $T$-{\em independent} -  $\xi=0$ 
and $d_o=$const. We plot $\rho(H=0)/\rho_*$, where $\rho=1/\sigma$ and
$\rho_* = 3 m \tau/(\pi e^2 N_F \ep_F)$. $s_\sigma(T)=1$ for all $T$ here.
The overall variation of  $\rho/\rho_*$ within a given $T$ range 
increases with $u$ and $g$. Hence, an appropriate choice of these 
parameters yields a resistivity which appears practically constant -
c.f. e.g. the experimental data in ref. \cite{klap} - although it has
the same shape as the curves of fig. 3.
Overall, the fit of eq. (\ref{sgma}) is good.
It can be seen that there is a discrepancy for low $T$.
At the moment, we cannot determine the precise origin of this effect.

In fig. 4 we plot the magnetoresistance corresponding to eq. (\ref{sgma}),
as a function of $(H/H_*)$.
The field at which $C_u=0$ is 
$H_*=\sqrt{|1-2 u +3u^2/4|/(1-u^2/4)}/(g_c \mu_B r)$, 
$g_c$ the gyromagnetic ratio of the carriers and $\mu_B$ the
Bohr magneton. 
Again, $s_\sigma(T)=1$ for all $T$.
The magnetoresistance for $H\rightarrow 0$ scales like $(H^2/T^\gamma)$, 
where $\gamma = 1.05$ for the parameter set (1) of fig. 3.
The typical range of values of $\gamma\simeq 1$ are in marked 
difference from $\gamma_W=2$ expected from the
conventional weak disorder magnetoresistance, but compare very favorably
with expts. \cite{magv}.

Turning to the conventional weak disorder conductivity contributions in
the presence of finite $M$,
it is easy to see that they can typically be written as
\be
\sigma_W = \frac{2 e^2}{m^2}\int d\vec{k}\; k_x^2 \frac{\Gamma^2 W}
{(1-\Gamma M)^2} \;\;,     \label{sm2}
\ee
in a manner analogous to eq. (\ref{sm1}) - but note the square in the 
denominator of eq. (\ref{sm2}).
$W$ stands for any such diagrammatic contribution, with or without 
interactions.
E.g. the conventional weak localization correction corresponds to
$W\equiv -C$, $C_{q,\om}$ being the Cooperon \cite{aa}.
From eqs. (\ref{sm1}) for $\sigma_M$ and (\ref{sm2}) for $\sigma_W$,
we obtain $\sigma_W \ll \sigma_M$ for $W \ll M$.
We believe this to be the explanation behind the negligible weak localization
etc. contributions observed in a number of experiments \cite{klap,coler}.
The same is true for the Hall coefficient as well, e.g. \cite{coler}. 
Note that, with increasing $H$, $M$ decreases, and the weak disorder
$\sigma_W$ is enhanced, as seen in \cite{coler}.
On the other hand, finite but small weak disorder contributions,
which behave like -ln($T$) in 2D \cite{aa,rev},
have also been observed in \cite{simm,senz}.
In the frame of our model, it is reasonable to expect that at
exponentially small $T$ $\sigma_W $ should dominate over $\sigma_M$,
and hence drive the system into the insulating regime.
In that case, we would {\em not} have a $T=0$ quantum metal-insulator phase
transition. However, there are two caveats here. Kopietz has shown that the 
quasiparticle weight vanishes for $T\rightarrow 0$ \cite{kopi}, 
in an interacting 2D Fermi gas with spinless disorder - the case 
with spin disorder remaining unresolved.
Moreover, the well known Kohn-Luttinger instability to a superconducting phase
may also interfere in the limit $T \rightarrow 0$. 
Hence the fate of the metallic state for $T \rightarrow 0$ is an open question.

Based on our model, we evaluate the specific heat $\delta C_V$ of the
carriers,
corresponding
to the same processes yielding $\sigma_M$ in eq. (\ref{sm1}). The free
energy is 
\be
F_M= N_F \om_o  \tau K \ln\Big(\frac{\ep_F-K\tau}{\ep_F+K\tau}\Big) s_F(T) \;,
\ee
with $K=\sqrt{g[z_1 F_1 + z_2 F_2]}$ 
and $\delta C_V=-T \partial^2 F_M/\partial T^2$. Here we took
into account the diagrams of figs. 1 and 2.
$s_F(T)$ has the same properties as $s_\sigma(T)$ above.
As before, this result should not be valid for $T\rightarrow 0$. We note
the pronounced $H$ dependence of $\delta C_V$.

To summarize, we have shown that interactions and spin disorder can drive 
the second order metal-insulator transition observed in 2D systems. 
The transition arises from the onset of strong spin-density correlations,
for a restricted range of the carrier density, the Zeeman energy and the 
temperature. We obtain good agreement with the experimentally
determined dependence of the conductivity on these parameters.

\vspace{.3cm}

I have enjoyed discussions and/or correspondence with P. Kopietz, 
P.T. Coleridge, A. Hamilton, E. Kiritsis,
S.V. Kravchenko, V. Senz, S. Vitkalov and D. Williams.
I acknowledge the hospitality of V.S. Vassiliadis in Cambridge.

\vspace{.5cm}

\centerline {\bf APPENDIX A}

\vspace{.4cm}

Equations (3), derived in \cite{gk},
are standard Bethe-Salpeter equations in
the spin-density channel, encompassing spin-scatering disorder and
interactions.
They are shown in fig. A1, and they
form a complete system, containing {\em all} possible 
combinations of the interaction $U$ and disorder lines in ladders,
i.e. in the "no-crossing lines" approximation. 

Their solution is given by
\be
A^1 = U (1-U {\cal D}^{-1})/D_o \;\;, 
\; A^0 = - U^2 {\cal D}^0/D_o \;\;, \;\;  \label{eaj}
\ee
where the determinant $D_o$ is 
\be
D_o = (1-U {\cal D}^{-1}) (1-U {\cal D}^{1}) - (U {\cal D}^0)^2 \;\;.\;\;
\ee

To obtain the solution, we substitute ${\cal D}^{i}$ according to the 
expressions given in the text. We perform the polynomial
expansions in $q^2, \om_m, \om_H$
in both numerator and denominator of the expressions resulting from 
eqs. (\ref{eaj}), after writing common denominators which are products of all
relevant denominators coming from the ${\cal D}^{i}$'s.
Then, taking the limit $Dq^2 < \om_m$, yields the solutions
in eq. (\ref{eqsol}). The solutions in the opposite limit
$Dq^2 > \om_m$ can be found in \cite{gk}. 

\vspace{.5cm}

\centerline {\bf APPENDIX B}

\vspace{.4cm}

In this Appendix we calculate the basic diagrammatic block B of fig. 1
\bea
B = T^4 \sum_{ \{\ep_1,\om_1,\ep_2,\om_2\} \{k_1,p_1,q_1,k_2,p_2,q_2\} }
G(k_1,\ep_1) G(k_1-q_1,\ep_1-\om_1) G(p_1,\ep_1) G(k-q_1,\ep+\Omega-\om_1) 
\\ \nonumber
G(k_2,\ep_2) G(k_2+q_2,\ep_2+\om_2) G(p_2,\ep_2) G(k+q_2,\ep+\om_2)
 \\  \nonumber
 A^j(q_1,\om_1) A^j(k'-k+q_1,\ep'-\ep+\om_1) \;\;
A^{j'}(q_2,\om_2)  A^{j'}(k''-k+q_2,\ep''-\ep+\om_2) \;\;  
D^{l,m_l}(k_1-k_2,\ep_1-\ep_2) \;\;.
\eea
Here the indices $l,m_l$ depend on $j$ and $j'$.
All the energies are Matsubara energies, but we supress the usual indices
$m,n$ for the moment.
The Green's function is 
$G(k,\ep_n) = 1/\{i\ep_n-\ep_k+i \text{sign}(\ep_n)/(2 \tau)\}$.
Fixing the external momenta and energies so that
\be
k'=k''=k \;\;,\;\; \ep'=\ep''=\ep \;\; \label{ekex},
\ee
the labels $k,\ep$ appear in both left and right external lines
of the diagram. Thus the diagram can be inserted in a conductivity bubble,
and calculate the conductivity accordingly. I.e. as a lowest order
approximation
- c.f. eq. (\ref{sm1}) for the full formula - B can contribute to the 
conductivity through the formula 
\be
\sigma_1 = \frac{2 e^2}{m^2}\int d\vec{k} \; k_x^2 
(G_R(k,\ep_F) G_A(k,\ep_F))^2 B \;\;.     \label{sm01}
\ee
Eq. (\ref{ekex}) implies directly that the two pairs of the interaction
propagators $A^j(q,\om)$ - corresponding to the upper and lower bubbles -
have identical arguments. This gives rise to the resonance factors
referred to in the text - c.f. the discussion around eq. (\ref{ew0}) 
and $R_j$ below.
Equivalently, taking identical arguments for each pair of $A^j(q,\om)$'s,
yields eq. (\ref{ekex}). 

In fact, the resonance in question {\em selects}
identical arguments $q,\om$ for each pair of $A^j(q,\om)$'s in diagrams of the
type of B above, in which this is possible. In general, one can construct
other diagrams, similar in structure to B, that is with pairs of 
$A^j(q,\om)$'s connected to a common bubble, but in these instances
it is {\em impossible} to obtain identical arguments $q,\om$ for the
two $A^j(q,\om)$'s. These diagrams present no special interest, as they
contain no resonance factors, and are ommitted.

Also, eq. (\ref{ekex}) implies that 
\be
p_1=k_1 \;\;,\;\; p_2=k_2 \;\;. 
\ee
Moreover, we can ommit the $q_1,\om_1,q_2,\om_2$ dependence of the 
two Green's functions in the external part of the two bubbles.

Taking all this into account we can write
\be
B = G_R(k,\ep) G_A(k,\ep) \; |R_j R_{j'}| \; \sum_q D^{l,m_l}(q,\om_T) \;\;,
\ee
where

\be
R_j = T^2 \sum_{0<\om_m, 0<\ep_n<\om_m, k, q} (A^j(q,\om_m))^2 \; G^2(k,\ep_n)
G(k-q, \ep_n-\om_m) \;\;, \label{rj1}
\ee
and we take $\om_T$ as the smallest non-zero Matsubara energy difference
\be
\om_T= 2 \pi T \;\;.
\ee
For the $q$ summations we take $Dq^2<\tau^{-1}$. Then,
\be
\sum_q D^{j,m_j}(q,\om_T) = \frac{F_{1,2}}{8 \pi^2 \tau^2 N_F D}  \;\;,
\label{logc}
\ee
where the functions $F_{1,2}$ appear after eq. (\ref{msi}) and $F_1$
corresponds to $D^{0,0}$, while $F_2$ to $D^{1,1}$ - c.f. eq. (\ref{ldif})
for $D^{j,m_j}$ .

\vspace{.5cm}

\centerline {\bf APPENDIX C}

\vspace{.4cm}

In this Appendix we evaluate the factor $R_j$.
We see that 
\be
R_j = T \sum_{\om_m, q} (A^j(q,\om_m))^2 F(q,\om_m) \;\;,
\ee
with 
\be
F(q,\om_m) = T  \sum_{k, 0<\ep_n<\om_m} G_R^2(k,\ep_n) G_A(k-q, \ep_n-\om_m)
\;\;.
\ee

We can ommit the $q$-dependence in $F(q,\om_m)$, $q<1/\sqrt{D\tau}$ being 
much smaller
than the relevant momenta of order $k_F$ in $G(k,\ep)$, and then we evaluate 
\bea 
I(k,\om_m) = T  \sum_{ 0<\ep_n<\om_m} G_R^2(k,\ep_n) G_A(k, \ep_n-\om_m)  \\
= \frac{1}{2\pi i} 
\int_{-\infty}^{\infty} d\ep \; n_F(\ep)\; [G_R^2(k,\ep) G_A(k, \ep-i \om_m)
- G_R^2(k,\ep+i \om_m ) G_A(k, \ep)] \;\;.
\nonumber
\eea
$n_F(\ep)$ is the Fermi-Dirac occupation function, and for low $T$ we have
\bea
2\pi i \; I(k,\om_m) \simeq  
\int_{-\infty}^{\ep_F} d\ep \; \Big\{\frac{1}{(\ep-\ek+it)^2}
\; \frac{1}{\ep-i\om_m-\ek-i t}
- \frac{1}{(\ep+i\om_m-\ek+it)^2}\; \frac{1}{\ep-\ek-i t} \Big\}  \\
= \frac{1}{\om_m +2t}\Big\{ \frac{1}{\ep_F+i\om_m-\ek+it}
-\frac{1}{\ep_F-\ek+it} + \frac{i}{\om_m +2t}
\ln\Big(\frac{(\ep_F-\ek)^2+t^2}{(\ep_F-\ek)^2+(\om_m+t)^2}\Big) \Big\} \;.
\nonumber
\eea
$t=1/(2\tau)$. Subsequently
\bea
F(\om_m) = N_F \int_{0}^{\infty} d\ek \; I(k,\om_m) \\
= \frac{1}{2\pi} \frac{1}{\om_m +2t} \Big\{\ln\Big(\frac{\ep_F+i\om_m+it}
{\ep_F+it}\Big) +\frac{i}{\om_m +2t} \Big[-2(\om_m +t)
\arctan\Big(\frac{\om_m +t}{\ep_F}\Big)+2 t \arctan\Big(\frac{t}{\ep_F}\Big)
\nonumber  \\ 
+ \ep_F \ln\Big(\frac{\ep_F^2+(\om_m+t)^2}{\ep_F^2+t^2}\Big) \Big] \Big\}
\nonumber \;\; .
\eea
In the limit
\be
\om_m \ll t \ll \ep_F \;\;
\ee
of interest here, we obtain
\be
F(\om_m) = \frac{ N_F \; \om_m}{4 \pi \ep_F^2} \;\;.
\ee

With $F(\om_m)$ in hand, we proceed to calculate
\bea
Y(\om_m)= \sum_{q} (A^j(q,\om_m))^2 = \frac{1}{4\pi D} \int_{0}^{\om_m} \;
(A^j(x=Dq^2,\om_m))^2 \; dx    \label{exyy} \\
= \frac{1}{4\pi D} \Big\{- \frac{1}{A_j^3}\; 
\frac{[A_u(K_{uj}+M_{uj}|\om_m|)-L_{uj}(C_u+B_u |\om_m|)]^2}
{C_u+B_u |\om_m| +A_u x} + Y_i \Big\} \Big|_{x=0}^{x=\om_m} \;\;, \nonumber
\eea
where we restrict ourselves to the dynamic limit $Dq^2 < \om_m$.
\be
Y_i = \frac{1}{A_u^3} \big\{ A_u L_{uj}^2 x +
2 L_{uj} [A_u(K_{uj}+M_{uj} |\om_m|)- L_{uj} (C_u+B_u |\om_m|)] \;
\ln(C_u+B_u |\om_m| +A_u x) \big\}\;\;,
\ee
is an irrelevant term, which does not yield a significant contribution
compared to the pole of $Y(\om_m)$ - despite the logarithmic singularity - and
will be ommited.

Subsequently we evaluate
\bea
R_j = T \sum_{\om_m>0} Y(\om_m) F(\om_m) = 
- T \sum_{\om_m>0} \frac{ F(\om_m)}{4\pi D A_u^3} \; 
\frac{[A_u(K_{uj}+M_{uj}\om_m)-L_{uj}(C_u+B_u \om_m)]^2}{C_u+B_u \om_m}
\label{expo} \\
\simeq \frac{T N_F \om_o \; (K_{uj}+M_{uj} \om_o)^2}
{16 i \pi^2 d \; \ep_F^2 A_u D} \;\;, 
\nonumber
\eea
where we approximated the $\om_m$ sum by the term most closely located to 
the pole $\om_m=\om_o$ - c.f. eq. (\ref{ew0}) - as $C_u<0$ and $B_u>0$,
and we ommited the term in eq. (\ref{exyy}) with $x=\om_m$, as it yields
a resonance at lower $\{T_o,\om_o\}$ than in eq. (\ref{ew0}).

\vspace{.5cm}
$^*$ Current address. E-mail : kast@iesl.forth.gr

\vspace{2cm}

\begin{figure}
\begin{center}
\epsfxsize8cm
\epsfbox{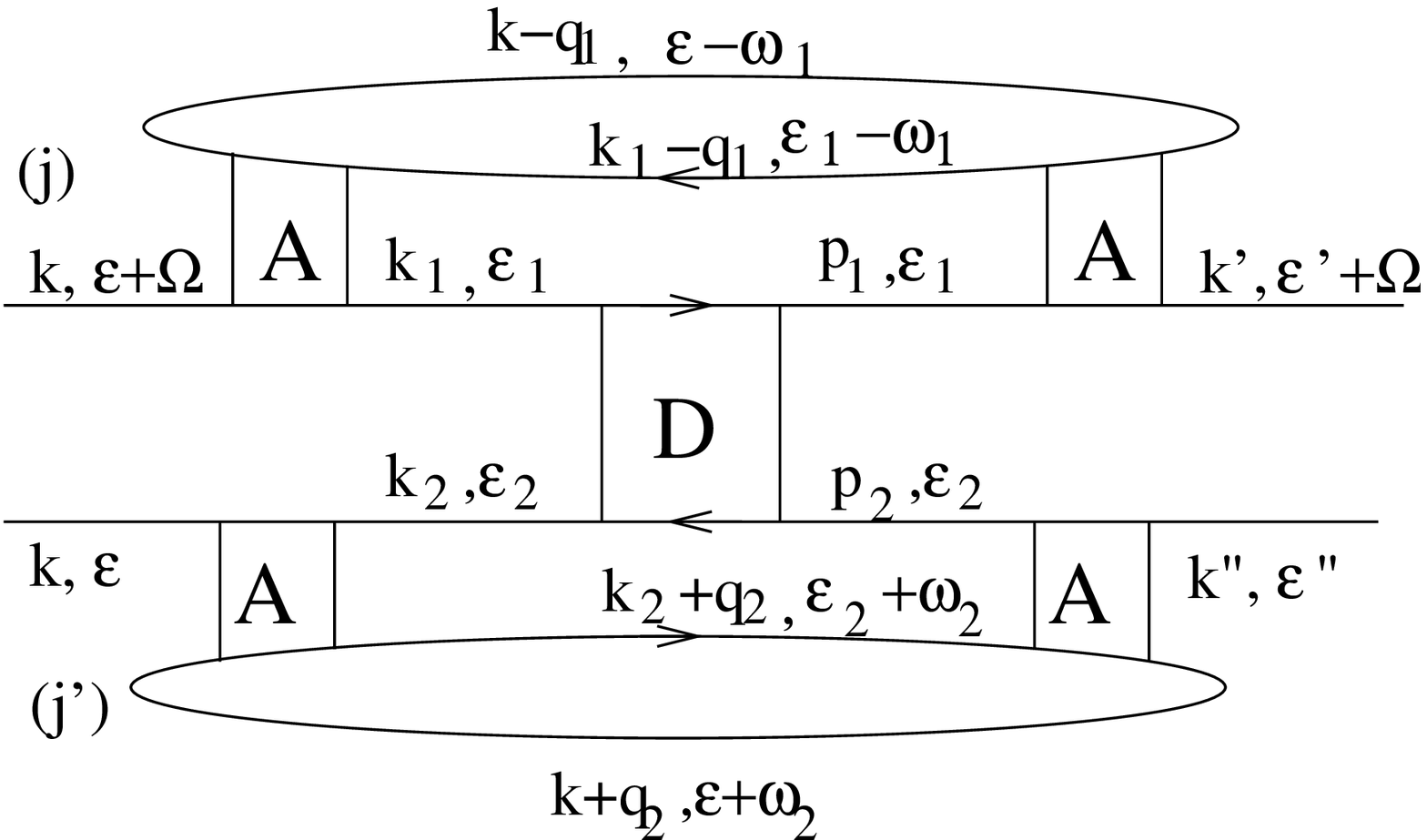}
\vspace{.1cm}
\centerline{Fig. 1}

The basic (lowest order) diagramatic block for $M_\sigma$.
$D_{q,\om}$ is the diffuson and A the renormalized
propagator of eq. (4).
All the appropriate spin structure diagrams are kept.
\end{center}
\end{figure}
 
\begin{figure}
\begin{center}
\epsfxsize7cm
\epsfbox{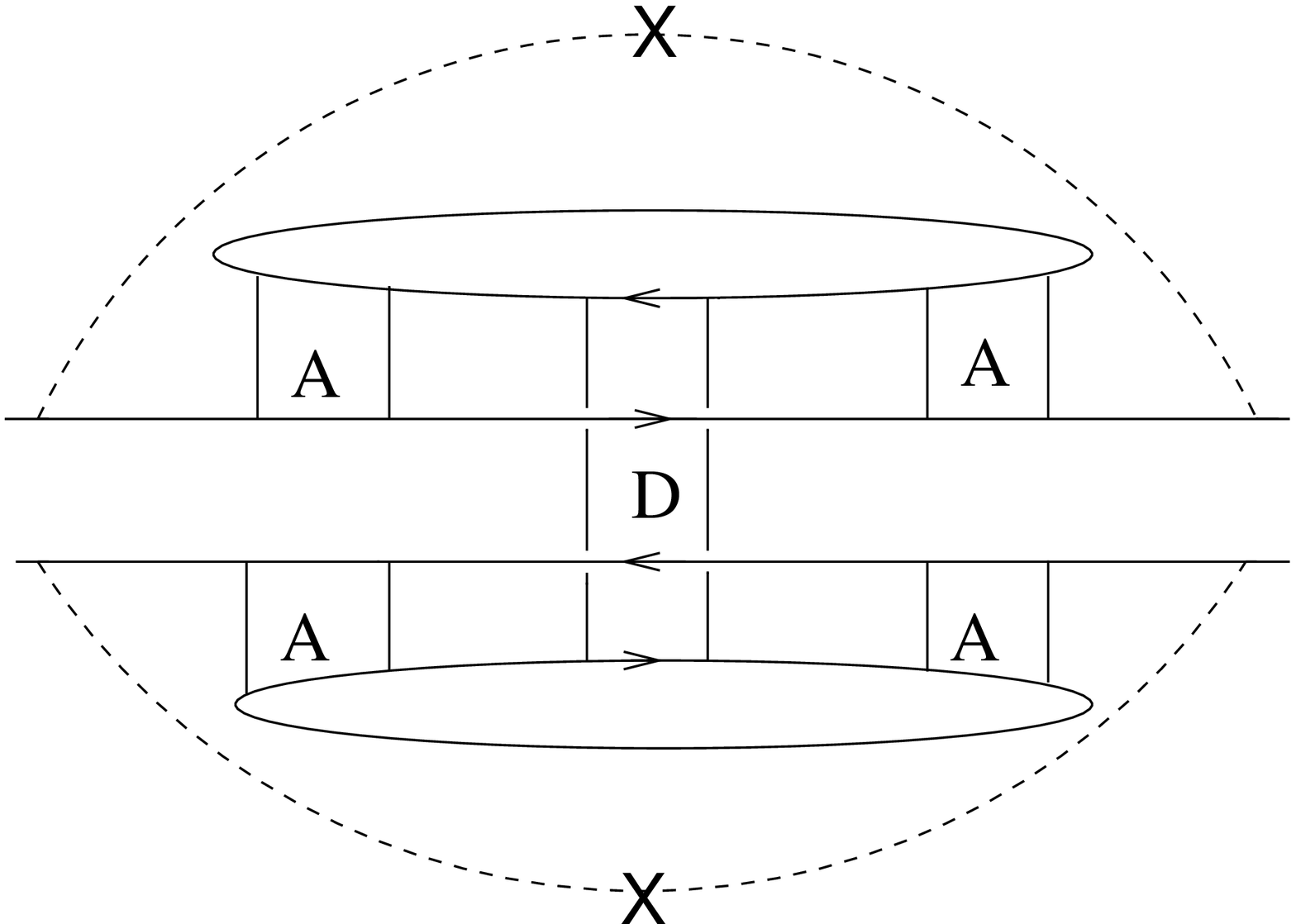}
\vspace{.1cm}
\centerline{Fig. 2}

Variations on the diagram of fig. 1. Here the diffuson connects the
bubble lines. One or two impurity scattering line(s) decorate the
main upper and/or lower line(s).
\end{center}
\end{figure}

\begin{figure}
\begin{center}
\epsfxsize10cm
\epsfbox{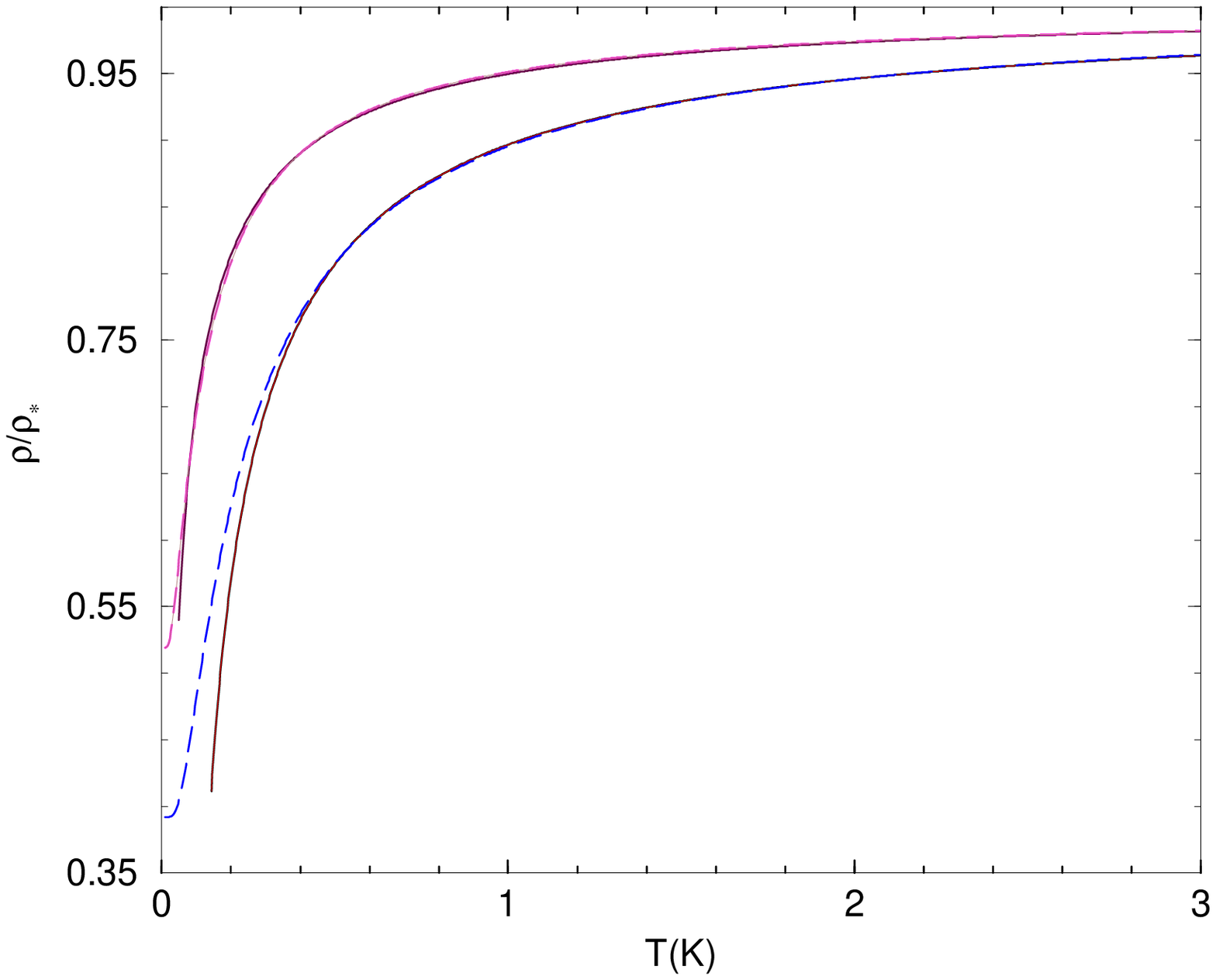}
\vspace{.1cm}
\centerline{Fig. 3}

Fit of eq. (\ref{sgma}) - continuous lines - to the experimentally 
determined resistivity form of eq. (\ref{ex0}) - dashed 
lines. Here $H=0$ and $\rho_* = 3 m \tau/(\pi e^2 N_F \ep_F)$.
$\chi=1$ and $\xi=0$. Both lower (1) and upper (2) curves have $u$=0.83,
$r=\tau^{-1}=2$ K, and only differ by the ratio $g^{(1)}/g^{(2)}=2$.
The parameters of eq. (\ref{ex0}) are $(\rho_o/\rho_*) = 0.392/0.519$,
$(\rho_1/\rho_*) = 0.61/0.48$, $T_*=0.192/0.1$ K and $k=1/0.98$, for  
curves (1) and (2) respectively.
\end{center}
\end{figure}

\begin{figure}
\begin{center}
\epsfxsize10cm
\epsfbox{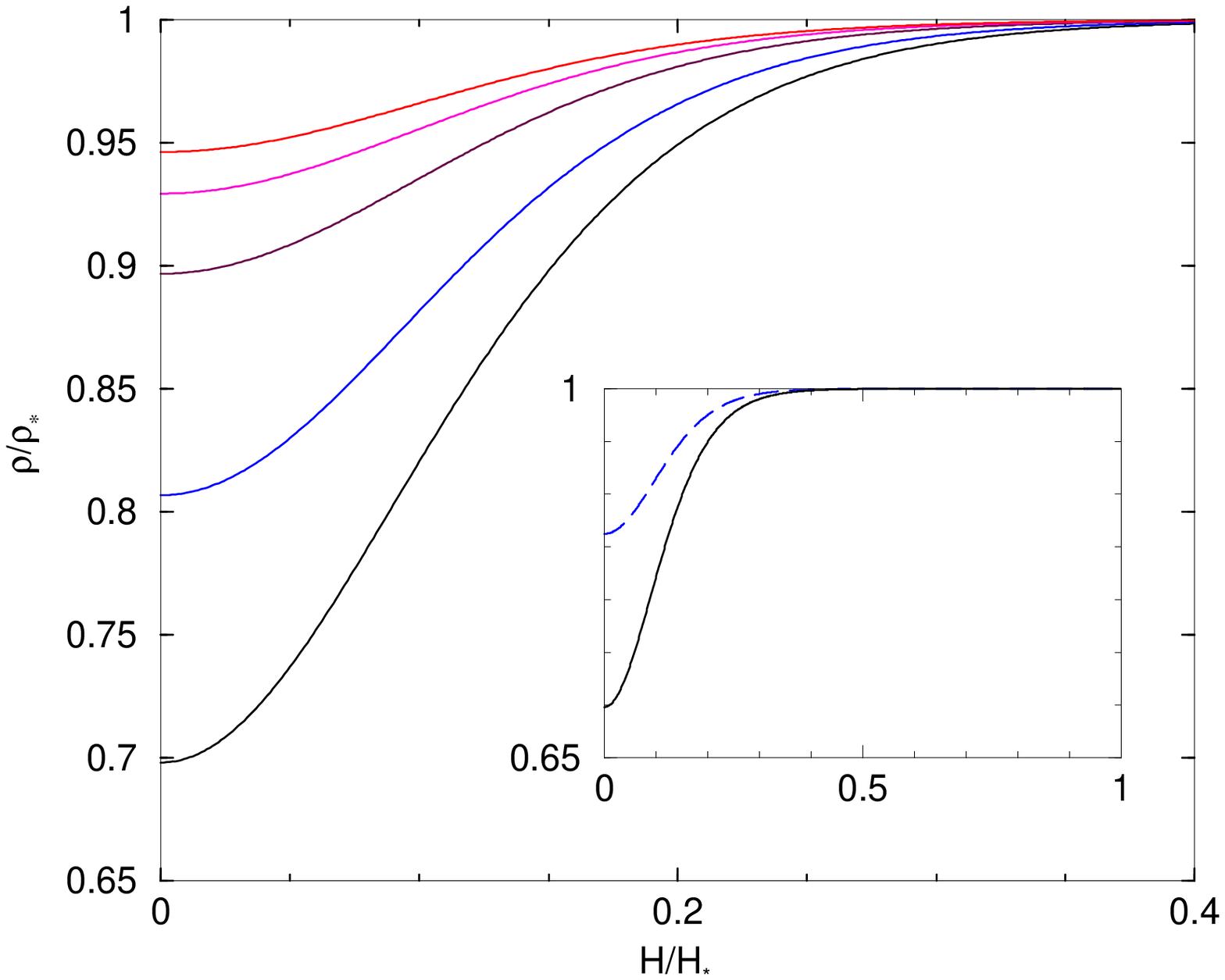}
\vspace{.1cm}
\centerline{Fig. 4}

Magnetoresistance from eq. (\ref{sgma}), for the parameter set (1) of 
fig. 3. $H_*$ is given in the text.
The curves bottom to top correspond to $T$=0.3, 0.5, 1, 1.5 and 2 K.
We note that $\rho/\rho_*=1$ implies $M=0$.
The full curves to $H=H_*$ for $T=0.3$ K are shown in the inset, the
dashed line corresponding to the parameter set (2) of fig. 3.

\end{center}
\end{figure}

\begin{figure}
\begin{center}
\epsfxsize8cm
\epsfbox{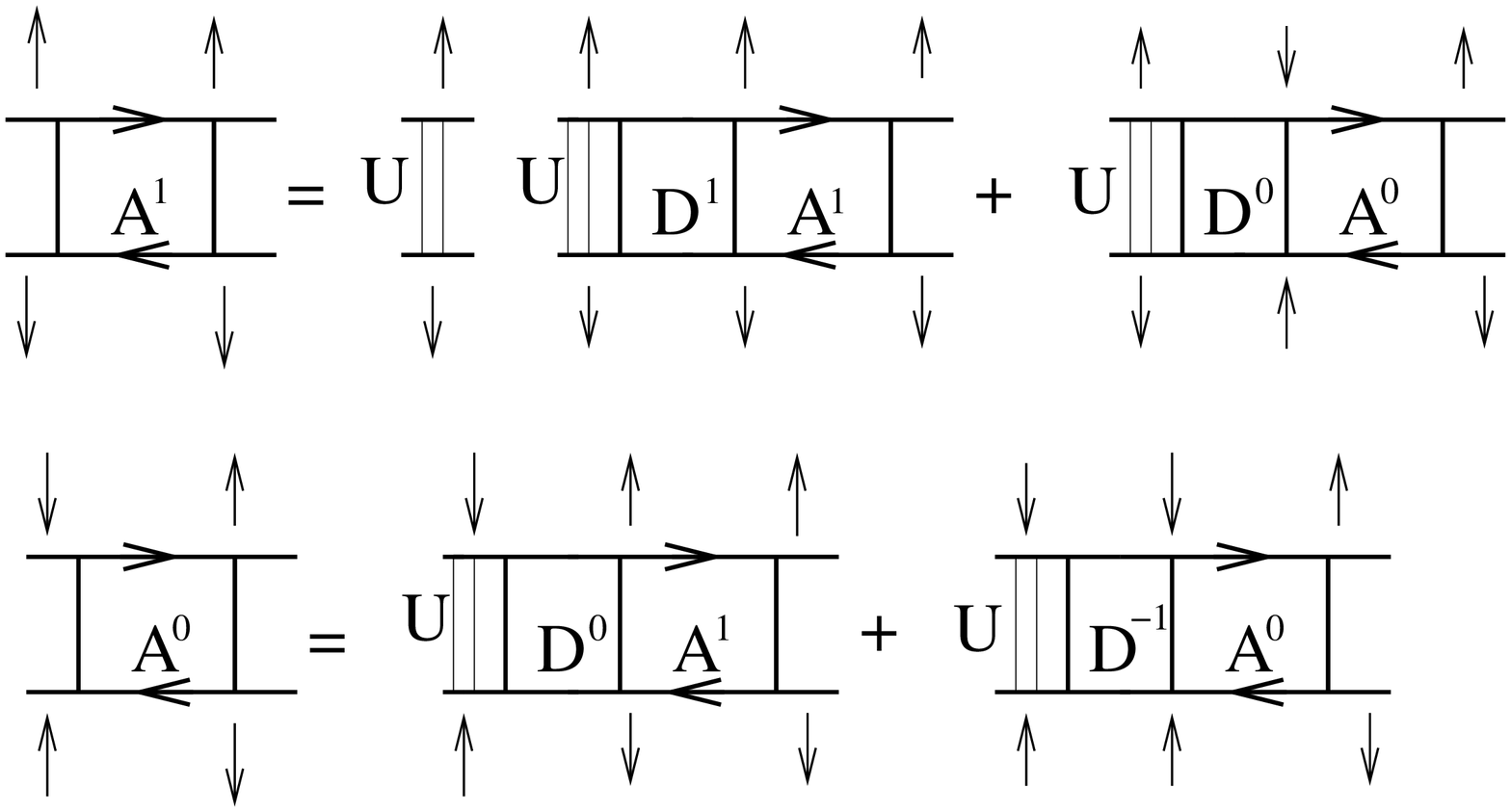}
\vspace{.1cm}
\centerline{Fig. A1}

The Bethe-Salpeter eqs. (3). The thin double line stands for $U$.
The factors ${\cal D}^i$ are given in the text. Note the explicit spin
indices of the various components of the diagrams.

\end{center}
\end{figure}


\begin{references}
\bibitem{prp}
The opposite limit $Dq^2>\om$ does not yield the effect sought for, despite
$B_u<0$ for $1< u < 1.2$.
\end{references}
\end{document}